\documentclass[twocolumn,showpacs,amsmath,amssymb]{revtex4}

\usepackage[T1]{fontenc}

\usepackage[latin1]{inputenc}

\usepackage{amssymb}

\usepackage{epsfig}

\usepackage{psfrag}
\usepackage{graphicx}% Include figure files
\usepackage{dcolumn}% Align table columns on decimal point
\usepackage{bm}% bold math

\begin{document}

\def\jddot#1{\stackrel{\bigdot\bigdot}{#1}}
\def\L{\mathcal L}
\def\e{\varepsilon}
\def\a{\alpha}
\def\b{\beta}
\def\c{\chi}
\def\d{\delta}
\def\e{\epsilon}
\def\f{\phi}
\def\g{\gamma}
\def\h{\eta}
\def\i{\iota}
\def\j{\psi}
\def\k{\kappa}
\def\l{\lambda}
\def\m{\mu}
\def\n{\nu}
\def\o{\omega}
\def\p{\pi}
\def\q{\theta}
\def\r{\rho}
\def\s{\sigma}
\def\t{\tau}
\def\u{\upsilon}
\def\x{\xi}
\def\z{\zeta}
\def\D{\Delta}
\def\F{\Phi}
\def\G{\Gamma}
\def\J{\Psi}
\def\L{\Lambda}
\def\O{\Omega}
\def\P{\Pi}
\def\Q{\Theta}
\def\S{\Sigma}
\def\U{\Upsilon}
\def\X{\Xi}

%Varletters

\def\ve{\varepsilon}
\def\vf{\varphi}
\def\vr{\varrho}
\def\vs{\varsigma}
\def\vq{\vartheta}

\def\dg{\dagger}                                     % hermitian conjugate
\def\ddg{\ddagger}                                   % double dagger
\def\wt#1{\widetilde{#1}}                    % big tilde
\def\mt{\widetilde{m}_1}
\def\mti{\widetilde{m}_i}
\def\mtDM{\widetilde{m}_{DM}}
\def\rt{\widetilde{r}_1}
\def\mtt{\widetilde{m}_2}
\def\mttt{\widetilde{m}_3}
\def\rtt{\widetilde{r}_2}
\def\mb{\overline{m}}
\def\VEV#1{\left\langle #1\right\rangle}        % < >
\def\be{\begin{equation}}
\def\ee{\end{equation}}
\def\ds{\displaystyle}
\def\ra{\rightarrow}

\def\bea{\begin{eqnarray}}
\def\eea{\end{eqnarray}}
\def\NO{\nonumber}
\def\Bar#1{\overline{#1}}

% Journal abbreviations (preprints)

\def\pl#1#2#3{Phys.~Lett.~{\bf B {#1}} ({#2}) #3}
\def\np#1#2#3{Nucl.~Phys.~{\bf B {#1}} ({#2}) #3}
\def\prl#1#2#3{Phys.~Rev.~Lett.~{\bf #1} ({#2}) #3}
\def\pr#1#2#3{Phys.~Rev.~{\bf D {#1}} ({#2}) #3}
\def\zp#1#2#3{Z.~Phys.~{\bf C {#1}} ({#2}) #3}
\def\cqg#1#2#3{Class.~and Quantum Grav.~{\bf {#1}} ({#2}) #3}
\def\cmp#1#2#3{Commun.~Math.~Phys.~{\bf {#1}} ({#2}) #3}
\def\jmp#1#2#3{J.~Math.~Phys.~{\bf {#1}} ({#2}) #3}
\def\ap#1#2#3{Ann.~of Phys.~{\bf {#1}} ({#2}) #3}
\def\prep#1#2#3{Phys.~Rep.~{\bf {#1}C} ({#2}) #3}
\def\ptp#1#2#3{Progr.~Theor.~Phys.~{\bf {#1}} ({#2}) #3}
\def\ijmp#1#2#3{Int.~J.~Mod.~Phys.~{\bf A {#1}} ({#2}) #3}
\def\mpl#1#2#3{Mod.~Phys.~Lett.~{\bf A {#1}} ({#2}) #3}
\def\nc#1#2#3{Nuovo Cim.~{\bf {#1}} ({#2}) #3}
\def\ibid#1#2#3{{\it ibid.}~{\bf {#1}} ({#2}) #3}

\title{Cold Dark Matter from heavy right-handed neutrino mixing}

\author{Alexey Anisimov}

\affiliation{Deutsches Elektronen-Synchrotron DESY, 22603 Hamburg, Germany}

\author{Pasquale Di Bari}

\affiliation{INFN, Sezione di Padova, Dipartimento di Fisica G.~Galilei, Via Marzolo 8, 35131 Padua, Italy}

\begin{abstract}

\noindent We show that, within the seesaw mechanism, an almost decoupled
right-handed (RH) neutrino species $N_{DM}$ with mass $M_{DM}\gtrsim 100\,{\rm GeV}$
can play the role of Dark Matter (DM). The $N_{DM}$'s  can be produced
from nonadiabatic conversions of thermalized (source) RH neutrinos
with mass $M_S$ lower than $M_{DM}$. This is possible if a
non-renormalizable operator is added to the minimal type I seesaw Lagrangian.
The observed DM abundance can be reproduced for
$M_{DM}\,\delta^{1/4} \sim 10^{-13}\,\Lambda_{\rm eff}\,\xi$,
where $\L_{\rm eff}$ is a very high energy new physics scale,
$\delta \equiv (M_{DM}-M_S)/M_{DM}$ and $\xi\lesssim 1$
is a parameter determined by the RH neutrino couplings.
\end{abstract}


\maketitle

\section{Introduction}

\vspace{-2mm}

The results from neutrino oscillation experiments
represent a success for the seesaw mechanism \cite{seesaw},
the simplest way to understand why neutrinos are massive,
yet so light compared to all other massive particles in
the Standard Model (SM).

Indeed, within the seesaw, the atmospheric and the solar neutrino mass
scales point to a high energy scale $\sim 10^{15}\,{\rm GeV}$
compatible with grand unification and at the same time one
can understand the observed large mixing angles.
Moreover, neutrino oscillations support leptogenesis \cite{fy},
an attractive way to explain the observed baryon asymmetry
of the Universe and a direct consequence of the seesaw mechanism.

Despite the great progress made in recent years in deriving,
especially from leptogenesis \cite{leptogenesis},
interesting constraints on those seesaw parameters that
escape the low energy experiments investigation,
we still lack a way to probe the seesaw mechanism.
The main obstacle is that, for natural choices of the seesaw parameters,
the heavy right-handed(RH) neutrinos, predicted by the seesaw, are not expected
to be detected at colliders, because they would be either too heavy or too weakly coupled.
Moreover they usually decay very fast disappearing from the cosmological lore.
If leptogenesis is the right explanation of the observed matter-antimatter asymmetry
of the Universe, produced from the $C\!P$ violating decays of the RH neutrinos,
this would be the only relic trace left over at present.

However, in this paper, we show that a weakly
coupled RH neutrino species can play the role of cold DM.
The scenario we present differs significantly from the one proposed in \cite{asaka},
where the lightest RH neutrino with a $\cal{O}({\rm KeV})$ mass plays the role of warm DM,
and neutrino Yukawa couplings are much smaller compared to charged
leptons and quarks Yukawa couplings.
In our model, we assume that all  RH neutrinos are {\em heavy}, with
the lightest RH neutrino mass not lower than the electroweak scale.
In this way, the neutrino Yukawa couplings can be of
the same order as for the other massive fermions.

\section{Failure of the minimal picture}

\vspace{-2mm}

The (type I) seesaw mechanism \cite{seesaw} is a minimal way
to explain neutrino masses. The SM Lagrangian is extended adding
a Yukawa interaction term between three RH neutrinos $\nu_R$
and the three left-handed doublets $l$ via a Higgs doublet $\phi$
and a Majorana mass term $M$,
\be\label{seesaw}
-{\cal L}_{Y+M}= \bar{l}_{L}\,\phi\,h\,\nu_{R}-{1\over 2}\,
\overline{\nu^c_{R}}\,M\,\nu_{R} + h.c. \, ,
\ee
where $h$ is the matrix of the neutrino Yukawa couplings.

After electroweak symmetry breaking, induced by the Higgs VEV $v$,
the Yukawa interaction generates a Dirac mass term $m_D=h\,v$.
In the seesaw limit, $M\gg m_D$, the spectrum of mass eigenstates
splits into three light neutrinos $\nu_i$ with masses given by the seesaw formula,
\be
{\rm diag}(m_1,m_2,m_3)= -U^{\dagger}\,m_D\,{1\over M}\,m_D^T\,U^{\star} \, ,
\ee
where $U$ is the leptonic mixing matrix,
and into three heavy neutrinos $N_i$ with masses
$M_1\leq M_2 \leq M_3$. These coincide, with very
good approximation, with  the eigenvalues of
the  Majorana mass matrix.

Neutrino oscillations experiments measure two
neutrino mass-squared differences. For normal schemes one has
$m^{\,2}_3-m_2^{\,2}=\Delta m^2_{\rm atm}$ and
$m^{\,2}_2-m_1^{\,2}=\Delta m^2_{\rm sol}$,
whereas for inverted schemes one has
$m^{\,2}_3-m_2^{\,2}=\Delta m^2_{\rm sol}$
and $m^{\,2}_2-m_1^{\,2}=\Delta m^2_{\rm atm}$.
For $m_1\gg m_{\rm atm} \equiv
\sqrt{\Delta m^2_{\rm atm}+\Delta m^2_{\rm sol}}=
(0.050\pm 0.001)\,{\rm eV}$ \cite{oscillations}
the spectrum is quasidegenerate, while for
$m_1\ll m_{\rm sol}\equiv \sqrt{\D m^2_{\rm sol}}
=(0.00875\pm 0.00012)\,{\rm eV}$ \cite{oscillations}
it is fully hierarchical (normal or inverted).
For definiteness we will refer to the case of normal
schemes but all the discussion applies to inverted schemes as well.

The RH neutrino decays can be conveniently described in terms of the decay parameters
$K_i \equiv \widetilde{\G}_{D i}/H(T=M_i)$, where $\widetilde{\G}_{D i}$
are the decay widths. These can be related to the neutrino masses
introducing the effective neutrino masses, defined as
$\widetilde{m}_{i}\equiv {(m^{\dagger}_D m_D)_{ii}/ M_i}$,
such that $K_i={\mti/ m_{\star}}$, where
$m_{\star}\simeq 1.08\times 10^{-3}\,{\rm eV}$.
Assuming $N_1$ to be heavier than the Higgs boson,

from the LEP bound \cite{LEP} one has $M_1 \gtrsim 115\,{\rm GeV}$
and the $N_i$ lifetimes are then given by
\be\label{tiKi}
\t_i= {8\,\pi\,v^2\over \mti\,M_i^2} \simeq {5\over K_i}\,
\left({{\rm TeV}\over M_i}\right)^2\times 10^{-13}\,{\rm sec} \, .
\ee
Let us now impose that one among the three RH neutrinos species $N_i$,
plays the role of DM particle which we indicate with $N_{DM}$.
This implies $\t_{DM}\geq t_0 \simeq 4\times 10^{17}\,{\rm sec}$,
where $t_0$ is the age of the Universe. However, since the
$N_{DM}$-decays would produce ordinary neutrinos, a much more
stringent lower bound comes from neutrino telescopes \cite{ellis},
\be\label{lb}
{\t_{DM}\over t_0} \gtrsim \a \gg 1 \, .
\ee
In the range $M_{DM}\sim 10^{5\div 9}$\,{\rm GeV},
the AMANDA limits on neutrino flux implies $\alpha\sim 10^9$
\cite{eglns,amanda}, while
in the range $M_{DM}\sim {10^{2\div 5}}\,{\rm GeV}$, where the
atmospheric neutrino flux is observed, the lower bound is more relaxed.
In any case, since strong future improvements are expected
from the ICE-CUBE experiment, we will leave indicated the dependence
on $\a$ in the following discussion \footnote{Constraints from $\gamma$-rays from
decaying Dark Matter in the Milky Way halo give similar or even stronger bound
depending on $M_{DM}$ \cite{bertone}.}.
From the relation (\ref{tiKi}), this translates into an
upper bound on the decay parameter $K_{DM}$ (or equivalently on
the effective neutrino mass $\widetilde{m}_{DM}$) given by
\be\label{lifetime}
K_{DM}\, (\widetilde{m}_{DM}/{\rm eV}) \lesssim {10^{-30 (33)}\over \a}\,
\,\left({\rm TeV \over M_i} \right)^2 \,.
\ee
Moreover, imposing that the $N_{DM}$-abundance
explains the measured DM contribution to the energy
density of the Universe, one finds a condition
on $r_{DM}\equiv (N_{N_{DM}}/N_{\gamma})_{\rm prod}$,
the ratio of the number of $N_{DM}$ to the photon number
at the time of the $N_{DM}$-production, occurring at temperatures
higher than the electroweak phase transition,
\be\label{prod}
r_{DM} \sim
10^{-9}\,(\O_{\rm DM}\,h^2)\,{{\rm TeV}\over M_{DM}}
\sim 10^{-10} \,{{\rm TeV}\over M_{DM}} \, .
\ee
Assuming that the correct value of $r_{DM}$ is produced by
some external mechanism, for example from inflaton decays,
a trivial DM model is obtained if the condition Eq.~(\ref{lifetime})
is satisfied. Within such a scenario one can indifferently identify either
$N_1$ or $N_2$ or $N_3$ with $N_{DM}$. The orthogonal seesaw matrix $\O$ \cite{casas},
is a useful tool to parametrize the Dirac mass matrix $m_D$, such that
\be\label{mD}
m_D=U\,D_m^{1/2}\,\Omega\,D_M^{1/2} \, ,
\ee
with $D_m\equiv {\rm diag}(m_1,m_2,m_3)$ and $D_M\equiv {\rm diag}(M_1,M_2,M_3)$.
The effective neutrino masses can then be
expressed as linear combinations of the neutrino masses
$\widetilde{m}_{i}=\sum_h\,m_h\,|\O_{hi}|^2$ and one
easily obtains $\mti\geq m_1$. Therefore,
the upper bound Eq.~(\ref{lifetime}) applies to $m_1$ as well,
implying hierarchical light neutrinos.
It also implies that $\O$
has to be close to the special form
\be\label{Omega}
\left(
\begin{array}{ccc}
  1  &  0   & 0   \\
  0  & \cos \o             & \sin \o \\
  0 &  -\sin \o & \cos \o
\end{array}
\right) \,\, ,
\ee
or to those other two obtained by column cyclic permutation.
Therefore, assuming exactly one of these three forms for the orthogonal matrix,
the condition Eq.~(\ref{prod}) is fulfilled only assuming
some mechanism for the $N_{DM}$-production based on physics
beyond the type I seesaw SM extension.
Even allowing small deviations from these special forms,
one undergoes a severe obstacle within the type I seesaw. Indeed one can think of
different processes producing the $N_{DM}$-abundance,
such as inverse decays or scatterings involving the top quark
or gauge bosons. However, in all cases one has approximately
$ r_{DM} \sim K_{DM}$ and it would then be impossible to
satisfy simultaneously the two requirements Eq.~(\ref{lifetime}) and Eq.~(\ref{prod}).

Let us consider a particular example that
clearly shows such a difficulty but that at the same time,
as we will see, will suggest a solution relying on a simple
and reasonable extension of the type I seesaw lagrangian.

We investigate the possibility that the
$N_{DM}$-production is induced by the mixing
of $N_{DM}$ with one of the other two RH neutrinos
acting as a source, and that we indicate with $N_{S}$.
Notice that $N_S$ has necessarily  a thermal abundance if
the reheat temperature is approximately higher than $M_S$.
This is because there cannot be more than one RH neutrino species
with $\mti \lesssim m_{\star}$.

For definiteness we can assume that $N_{DM}$ and $N_S$
are the two lightest RH neutrinos and hence there
are only two possibilities: either $M_{DM}=M_1$ and $M_{S}=M_2$
or vice-versa. In this case $N_3$ does not play any role
in the $N_{DM}$-production but it is necessary to
reproduce correctly the neutrino masses.

This scenario is realized choosing the following form for
the orthogonal matrix
\be\label{form}
\O'=\left(
\begin{array}{ccc}
 \sqrt{1-{\ve}^2}  &  -{\ve}         & 0 \\
            {\ve}  & \sqrt{1-{\ve}^2} & 0 \\
  0 & 0 & 1
\end{array}
\right) \,\, ,
\ee
representing a perturbation, with $\cos\o=1$, of the special
form in Eq.~(\ref{Omega}).
Here the prime index indicates that we are reexpressing $\O$ into
a basis where the RH neutrino mass term is still diagonal but
in a way that $M_{DM}$ is always the first eigenvalue and $M_{S}$
the second eigenvalue. Notice that we can choose $\ve$ real
and for convenience positive. Moreover notice that the choice
$\cos\o= 1$ is not restrictive. Indeed, in any case a value
$\cos\o\neq 1$ would not be relevant  for the DM production but notice that
it would be important if one simultaneously imposes successful leptogenesis
from $N_S$ decays, a possibility that will be discussed elsewhere \cite{prep}.

In order to describe the RH neutrino mixing, it is convenient to work
in the ``Yukawa basis'', where the Yukawa interaction term is diagonal.
This can be diagonalized by mean of a bi-unitary transformation,
$D_h\equiv {\rm diag}(h_A,h_B,h_C)=V^{\dagger}_L\,h\,U_R$.
The RH neutrino mixing matrix $U_R$ can be
found considering that it diagonalizes $h^{\dagger}\,h$, namely
$U^{\dagger}_R\,(h^{\dagger}\,h)\,U_R={\rm diag}(h^{\,2}_A,h^{\, 2}_B,h^{\, 2}_C)$.
Then, from the expression Eq.~(\ref{mD}), one can see that
our choice for $\O'$ simply results into

\be
U_R= \left(
\begin{array}{ccc}
\cos\theta & -\sin\theta & 0\\
\sin\theta & \cos\theta & 0 \\
0 & 0 & 1
\end{array} \right) \, ,
\ee
with $\sin\theta\simeq \ve\,\sqrt{M_S/M_{DM}}$ and into
\be\label{eigen}
h_{A}\simeq {\sqrt{m_1\,M_{DM}}\over v}\, ,
h_{B}\simeq {\sqrt{m_{\rm sol}\,M_S}\over v} \, ,
h_{C}\simeq {\sqrt{m_{\rm atm}\,M_3}\over v} \, .
\ee

This clearly shows that though $N_3$ does not
mix, it is necessary to reproduce the atmospheric
neutrino mass scale. Imposing the condition (\ref{lifetime}),
one can see that $\ve$ has to be tiny. Indeed one has
\be
\mt\simeq m_1 +m_{\rm sol}\,|\ve|^2  \, ,
\ee
and therefore the upper bound Eq.~(\ref{lifetime}) translates into the upper bounds
\footnote{In the exact limit $m_1=0$ the eigenvalue $h_A=0$. Moreover, in this limit,
plugging $\O'=I$ (cf. (Eq.~\ref{form})) into the Eq.~(\ref{mD}), one can immediately
see that the Yukawa coupling matrix, in the basis where the
RH neutrino mass matrix is diagonal, contains a vanishing column corresponding to the
DM RH neutrino. This is quite an obvious result, since it corresponds to impose
$\tau_{DM}=0$ that implies $h_{\alpha DM}=0$.
Therefore, the model corresponds to a special textured
form for the Yukawa couplings equivalent
to impose a particular symmetry such that the lagrangian is
invariant under a proper transformation of the neutrino fields.}
\be\label{eps}
{m_1\over \rm eV} \lesssim {10^{-33}\over \a}\,
\,\left({\rm TeV \over M_{DM}} \right)^2 \, , \hspace{4mm}
|\ve| \lesssim {10^{-16}\over \sqrt{\a}}\,
\left({{\rm TeV} \over M_{DM}}\right) \, .
\ee
This implies a hierarchical light neutrino spectrum and a
tiny mixing angle between the two lightest RH neutrinos.
The description of the production of the $N_{DM}$-abundance proceeds very
similarly to the case of light active-sterile neutrino oscillations
\cite{light} and in particular to the case described in \cite{zener},
where transitions occur in the non-adiabatic regime as it will
prove to be in our case. Let us write down the hamiltonian for the two
lightest RH neutrinos in the Yukawa basis.
This will be the sum of two terms: a pure kinetic term
and a second term accounting for matter effects described by
a potential that in the Yukawa basis is diagonal and given by \cite{weldon}
\be\label{V}
V_I \sim h^2_I\,T^2/(8\,k) \;\;\;\;\;\;\;\;  (I=A,B) \, ,
\ee
in the approximation of ultrarelativistic neutrinos,
implying $E\sim k$ and  $T\gg M_S/3$.
Notice that in any case for $T\lesssim M_S$ the $N_S$-abundance
is exponentially suppressed and the $N_{DM}$-production would stop anyway.
In order to further simplify the problem, we also employ a monochromatic approximation
where all neutrinos have the same mean energy value $k \sim 3 \,T$.
As usual, we can subtract from the hamiltonian a term proportional to the
identity, irrelevant in neutrino oscillations. Therefore, in the Yukawa basis,
the  relevant hamiltonian can be recast as
\be
\D H=
{\D M^2\over 12\,T} \left(\begin{array}{cc}
-\cos 2\theta+(v_A-v_B) & \sin 2\theta \\
 \sin 2\theta  & \cos 2\theta-(v_A-v_B) \,
\end{array}\right) \, ,
\ee
where we defined $v_I \equiv T^2\,h^2_I / (4\,\Delta M^2)$
and $\D M^2 \equiv M^2_{S}-M^2_{DM}$.
Approximating $\cos 2\theta\simeq 1$,
one can see that there is a resonance at a temperature
\be\label{Tres}
 T_{\rm res} \simeq
2\, \sqrt{\Delta M^2\, \over h^2_A-h^2_B }\simeq 2\, {\sqrt{- \Delta M^2}\, \over h_B } \, ,
\ee
only if $\D M^2<0$, i.e. only if  $M_1=M_{S} < M_{DM}=M_2$.
Using the Eq.~(\ref{eigen}), $T_{\rm res}$ can be conveniently recast as
\be\label{Tres2}
T_{\rm res} \simeq 10^7\,M_{DM}\,\sqrt{{v\over M_S}
\,\left(1-{M_S^2\over M_{DM}^2}\right)} \, .
\ee
If $M_{DM}\gtrsim 2\,M_S$ one has
$T_{\rm res}\simeq 10^7\,M_{DM}\,\sqrt{v/M_S}$. In this case, introducing
$z_{\rm res}\equiv M_{DM}/T_{\rm res}\simeq 10^{-7}\,\sqrt{M_S/v}$, one can
envisage a problem. The $N_S$'s thermalize  for
$z_{\rm eq}\simeq (6/K_S)^{1/3}\simeq 0.8$ \cite{pedestrians}.
Imposing $z_{\rm res}>z_{\rm eq}$ leads to an unacceptably
large values of $M_S, M_D$ and of the reheat temperature. Therefore, unless one
assumes an initial thermal abundance, one is forced to consider the degenerate
limit, for $\d\equiv (M_{DM}-M_{S})/M_{DM} \ll 1$. In this limit one
now obtains $T_{\rm res}\simeq 10^7\,M_{DM}\,\d^{1/2}\,\sqrt{v/M_{DM}}$
and $z_{\rm res}\simeq 10^{-7}\,\d^{-1/2}\,\sqrt{M_{DM}/v}$. For
$\d\lesssim 10^{-13}\,M_{DM}/{\rm TeV}$,
this time one can have $z_{\rm res}\gtrsim z_{\rm eq}$.
Therefore, the degenerate limit has to be considered
as a more attractive option.

Because of the tiny mixing angle the transitions
at the resonance occur in the nonadiabatic regime.
Indeed let us calculate the adiabaticity parameter at the resonance,
\be
\left.\gamma_{\rm res}\equiv {1\over 2\,\dot{\theta}_m\,{\ell}_m}\right|_{\rm res}=
\sin^2 2\theta\,{|\Delta M^2|\over 6\,T_{\rm res}\,H_{\rm res}} \, .
\ee
Here $H_{\rm res}\simeq 1.66\,\sqrt{g_{\star}}\,T^2_{\rm res}/M_{\rm Pl}$
is the value of the expansion rate at the resonance. Using the conditions
Eq.~(\ref{eps}) and Eq.~(\ref{lb}), one obtains the upper bound
$\gamma_{\rm res}\lesssim 10^{-26}\,\left({{\rm TeV} / M_{DM}}\right)^2$ .
The $N_{DM}$-abundance $r_{DM}$ can then
be calculated as the fraction of $N_S$'s that is converted into
$N_{DM}$. This is approximately given by the Landau-Zener formula,
\be
r_{N_{DM}}\sim {N_{DM}\over N_{S}}\sim
(1-e^{-{\pi\over 2}\,\gamma_{\rm res}}) \simeq {\pi \over 2}\,\gamma_{\rm res} \, .
\ee
Comparing with the condition Eq.~(\ref{prod}), it is evident that
neutrino mixing between heavy RH neutrinos cannot produce the
right $N_{DM}$ abundance, at least not within a minimal
type I seesaw extension of the SM. This conclusion is
confirmed by more precise calculations
beyond the Landau-Zener approximation.

\section{A way-out from nonrenormalizable operators}

Let us consider the possibility that adding
higher dimensional effective operators
to the minimal type I seesaw Lagrangian Eq.~(\ref{seesaw}),
while not affecting neutrino masses and mixing, enhances the
$N_{DM}$-production from neutrino mixing. In particular let us
consider the following dim-five effective operator
\footnote{The idea that this operator could enhance RH neutrino
oscillations was presented preliminarily  in \cite{proc}. It has then been also
considered for non-resonant production of the sterile DM neutrinos in \cite{bg}.}
\be
{\mathcal L}_{\rm eff}\propto {\lambda_{AB}\over \Lambda_{\rm eff}}|\Phi|^2\bar N_A^cN_B \, ,
\label{eff}
\ee
where $\Phi$ is the usual Higgs field, $\lambda$ is
a dimensionless coupling matrix and $\L_{\rm eff}$ is an unspecified
very high energy new physics scale that we treat as a free parameter.

This operator yields a new contribution to `matter effects'
into the hamiltonian \cite{prep}, that in the Yukawa basis can be written as
\be\label{Heff}
H_{\rm eff} \simeq {T^2\over 12\,\L_{\rm eff}}\,\l_{IJ} \, .
\ee
This result follows from the computation of the temperature dependent finite
real part of the RH neutrino self-energy \cite{weldon}:
\be
{\rm Re}[\Sigma_N(T)]={\lambda_{IJ}\over\Lambda}\int {d^4q\over (2\pi)^3}\delta(q^2-m^2_{\Phi})n_b(q),
\ee
where $n_b(q)={1\over e^{|q\cdot u|}-1}$ is the Bose-Einstein distribution with $u$ being the four-velocity of the thermal bath.
Assuming zero Higgs mass one then immediately deduce corresponding correction
to the Hamiltonian  (\ref{Heff}).

We can reasonably assume that $h^2_B \gg T_{\rm res}/\L_{\rm eff}$.
In this way in the Yukawa basis the total interaction term is
approximately still diagonal and with the same eigenvalues.
The relevant hamiltonian describing neutrino oscillations
can then be written as
\be
\D H^{\rm eff} \simeq
{\D M^2\over 12\,T} \left(\begin{array}{cc}
-v_B & \sin 2\theta+v_{\rm eff}^{AB} \\
 \sin 2\theta+v_{\rm eff}^{AB}  & v_B \,
\end{array}\right) \, ,
\ee
where we introduced
$v_{\rm eff}^{IJ}\equiv T^3\,\l_{IJ}/(2\, \D M^2\,\L_{\rm eff})$.
 Notice that the resonance condition
on the temperature, Eq.~(\ref{Tres}), does not change.
However, now the mixing angle is different and receives
a contribution from the off-diagonal terms in $H_{\rm eff}$,
such that $\sin 2\theta_{\rm eff} \simeq v_{\rm eff}^{AB}$ \,.

Imposing again that mixing is responsible for the DM production,
since we know that the mixing angle $\theta$ induced by the
Yukawa coupling $h_A$ is by far too small to play any role,
it can be assumed to be exactly zero.
This is a good feature since otherwise one could have objected
that radiative corrections could induce a large value anyway,
spoiling the stability of $N_{DM}$ on cosmological scales.
However, if it is exactly zero, one can invoke some symmetry
that protects it from radiative corrections.

Therefore, the adiabaticity parameter can now be  written as
\be
\gamma_{\rm res}^{\rm eff} \simeq
\sin^2 2\theta_{\rm eff}\,{|\D M^2|\over 6\,T_{\rm res}\,H_{\rm res}}
\simeq {\sqrt{|\D M^2|}\,M_{\rm Pl}\over 5\,\L_{\rm eff}^2\,\xi^2}\, ,
\ee

where we used the Eq.~(\ref{Tres}) for $T_{\rm res}$ and defined
$\xi\equiv g_{\star}^{1/4}\,h_B^{3/2}/\l_{AB}$.
Using again the Landau-Zener approximation for an estimation of
the $N_{DM}$ abundance, $r_{N_{DM}}\sim \gamma_{\rm res}$,
and imposing again the condition Eq.~(\ref{prod}),
we obtain the condition
\be\label{master}
M_{DM}\,\delta^{1\over 4} \sim 10^{-13}\,\L_{\rm eff}\,\,\xi \, .
\ee
It is easy to verify that the assumption $h^2_B \gg T_{\rm res}/\L_{\rm eff}$,
translates into a condition
$M_S\gg 10^{-2}\,{\rm GeV}\,g_{\star}^{1/3}\,\delta^{2/3}/\l_{AB}^{4/3}$ ,
easily verified except for tiny values of $\l_{AB}$.
Notice also that using the Eq.~(\ref{eigen}),
one can recast $\xi\sim (10^{-9}/\l_{AB})\,(M_S/{\rm TeV})^{3/4}$.
From the condition Eq.~(\ref{master}), one then finds
in   the hierarchical case, i.e. $M_{DM}\gtrsim 2\,M_S$,
\be
M_{S} \lesssim \left(\L_{\rm eff}\over 10^{13}\,{\rm TeV}\right)^4 \,
               \left(10^{-9}\over \l_{AB} \right)^4 \,{\rm TeV} \, ,
\ee
showing that in order not to satisfy $M_S\gtrsim 100\,{\rm GeV}$
the couplings cannot be too large. On the other hand in the more
interesting degenerate limit ($\d \ll 1$) one finds
\be
M_{DM} \gg \left(\L_{\rm eff}\over 10^{13}\,{\rm TeV}\right)^4 \,
               \left(10^{-9}\over \l_{AB} \right)^4 \,{\rm TeV} \, ,
\ee
showing, conversely, that in order not to have too large values of
$M_{DM}$ the couplings cannot be too small. Notice that too large values
$\log (M_{\rm DM}/{\rm TeV})\lesssim 5\div 8$ would spoil the cosmologically
stability of $N_{DM}$, leading to unobserved neutrino fluxes
at neutrino telescopes. Indeed in this case the nonrenormalizable operator
and the mixing with $M_S$ would induce too fast decays of the $N_{DM}$'s into
Higgs and leptons
\footnote{This can be also regarded saying that the
non-renormalizable operator breaks the symmetry that brings the Yukawa
coupling matrix to the
special textured form which is  needed
to have $N_{DM}$ cosmologically stable.}.
For $\L_{\rm eff}\sim M_{GUT}\div M_{Pl}$
one has then $\l_{AB}\gtrsim 10^{-13\div -10}$. The smallness of
$\lambda_{AB}$ can be explained in two ways. In the case when $\L_{\rm eff}\sim M_{GUT}$
the operator (\ref{eff}) can be generated radiatively from the coupling
to the GUT scale particles. For example, one can assume the Yukawa coupling (with the strength $h$)
between RH neutrino, Higgs and heavy $(m\sim M_{GUT})$ fermion. This coupling generates at one loop
the operator (\ref{eff}) after heavy fermion is integrated out.
The values of $\lambda_{AB}$ are, therefore, given by $h^2(T_{\rm res})$ and,
if  $h(T_{\rm res})\gtrsim 10^{-4\div -5}$, they come out naturally in the desired region.
Alternatively, if the operator (\ref{eff}) is generated gravitationally ($\L_{\rm eff}\sim M_{Pl}$) the smallness of the coefficients $\lambda_{AB}$ can be explained in the models where the effective value of
$M_{Pl}$ in the early universe is different from its present value (e.g. see Refs.\cite{xiphi}). However, the consequent decay channels at present should be estimated with $\lambda_{AB}\sim 1$.
 A detailed analysis of the constraints from decays will be presented
elsewhere \cite{prep}, however, it is remarkable that
the mechanism is viable for reasonable values of the involved parameters.

\section{Conclusions}

We presented a new scenario where the role of DM is played by
heavy RH neutrinos. The scenario is based on a mechanism where
the DM RH neutrinos are produced through mixing enhanced by the
additional presence of higher dimensional effective operators
into the usual type I seesaw Lagrangian. The mechanism relies crucially
on the fact that is necessary to convert just a very small fraction of
the source RH neutrinos into the DM RH neutrinos.
In this way the additional operator has the effect of enhancing
the mixing without spoiling any other successful feature of the
type I seesaw mechanism and at the same time preserving the DM RH neutrinos
stability on cosmological times. A straightforward prediction of the
mechanism is that the lightest neutrino mass has to vanish.
It also seems quite general that the DM RH neutrinos decay and this could lead
to signatures in cosmic rays.  The recent detected excess of positrons
in the HEAT and PAMELA experiments have been interpreted
as due to decaying DM particles with a mass higher than $300\,{\rm GeV}$
and a lifetime of approximately $\tau_{DM}\sim 10^{26}\,{\rm sec}$ \cite{ibarra}.
Therefore, our mechanism seems
to have the right features to explain this excess.
These results are quite interesting since not only they are fully compatible
with our model but also because the value for the life time corresponds to
the saturation of the lower bound Eq.~\ref{lb} from the AMANDA data when
$M_{DM}\sim 10^{5\div 9}~{\rm GeV}$ and a signal should be expected
from the ICE CUBE experiment.

It should also be noticed that the special orthogonal form Eq.~(\ref{form})
predicted by the mechanism corresponds \cite{geometry} to a particular
sequential dominated model \cite{king}. Therefore, the proposed
scenario for the solution of the DM conundrum restricts remarkably the
seesaw parameter space, providing a potential smoking gun for
the seesaw mechanism.

\vspace{2mm}

We wish to thank F.~Bezrukov, S.~Davidson, F.~Feruglio, A.~Ibarra, M.~Losada,
E.~Nardi, S.~King, A.~ Riotto and  S.~Sibiryakov for useful discussions.
PDB is supported by the Helmholtz Association of
National Research Centres, under project VH-NG-006.

\end{document}